\documentclass{revtex4}
\usepackage{amssymb}
\usepackage{amsfonts}
\usepackage{amsmath}

\input{tcilatex}

\begin{document}

\title{Position-dependent mass harmonic oscillator: classical-quantum
mechanical correspondence and ordering-ambiguity}
\author{Omar Mustafa}
\email{omar.mustafa@emu.edu.tr}
\affiliation{Department of Physics, Eastern Mediterranean University, G. Magusa, north
Cyprus, Mersin 10 - Turkey,\\
Tel.: +90 392 6301078; fax: +90 3692 365 1604.}

\begin{abstract}
We recycle Cruz et al.'s (Phys. Lett. A 369 (2007) 400) work on the
classical and quantum position-dependent mass (PDM) oscillators. To
elaborate on the ordering ambiguity, we properly amend some of the results
reported in their work and discuss the classical and quantum mechanical
correspondence for the PDM harmonic oscillators. We use a point canonical
transformation and show that one \textit{unique} quantum PDM oscillator
Hamiltonian (consequently, one \textit{unique} ordering-ambiguity parametric
set $j=l=-1/4$ and $k=-1/2$) is obtained. To show that such a parametric set
is not just a manifestation of the quantum PDM oscillator Hamiltonian, we
consider the classical and quantum mechanical correspondence for quasi-free
PDM particles moving under the influence of their own PDM force fields.

\end{abstract}

\maketitle

\section{Introduction}

The assumption that the information on the material properties is encoded in
the mass of the non-relativistic quantum particle suggests the well known
position-dependent mass (PDM) von Roos Hamiltonian (cf. e.g., \cite{1})%
\begin{equation}
\hat{H}=\frac{1}{4}\left[ m\left( x\right) ^{\boldsymbol{j}}p_{x}m\left(
x\right) ^{k}p_{x}m\left( x\right) ^{l}+m\left( x\right) ^{l}p_{x}m\left(
x\right) ^{k}p_{x}m\left( x\right) ^{j}\right] +V\left( x\right) ,
\end{equation}%
where the ordering-ambiguity parameters $j$, $k$, and $l$ are subjected to\
the von Roos constraint $j+k+l=-1$. Changing the values of $j$, $k$, and $l$
would change the profile of the kinetic energy term (hence changing the
profile of the effective potential) and therefore an ordering ambiguity
conflict emerges in the process. This has inspired many research
contributions over the last few decades, some of which were developed from
theoretical points of view \cite{2,3,4,5} and others were developed to
generate exactly solvable problems \cite{6,7,8,9,10,11,12,13,14,15,16,17}.
On the theoretically acceptable sides, nevertheless, it is found that the
continuity conditions at the abrupt heterojunction between two crystals
enforce the parametric condition $j=l$ (cf., e.g., Koc et al. \cite{4}). A
condition that makes the parametric proposals of Ben Daniel and Duke ($%
j=l=0, $ $k=-1$), Zhu and Kroemer ($j=l=-1/2,$ $k=0$), and Mustafa and
Mazharimousavi ($j=l=-1/4,$ $k=-1/2$) (cf., e.g., \cite{3,4} ) survive among
others available in the literature.

Very recently, on the other hand, Mazharimousevi and Mustafa \cite{3} have
argued that the fixation of the ordering-ambiguity parameters may be sought
through the classical observations of a given free PDM-particle moving under
the influence of its own internally byproducted force field. They have
observed that Zhu and Kroemer's \cite{4} ($j=l=-1/2$, $k=0$), and Mustafa
and Mazharimousavi's \cite{4} ($j=l=-1/4$, $k=-1/2$) orderings have provided
consistent quantum mechanical correspondence to the classical observations.
They have also suggested that the Gora and William's \cite{4} ($k=l=0$, $%
j=-1 $) and Li and Kuhn's \cite{4}\ ($k=l=-1/2$, $j=0$) orderings should be
disqualified, not only on the grounds of the continuity conditions at the
abrupt heterojunction but also on the grounds of failing to provide a
consistent quantum correspondence to classical observations. Such classical
and quantum correspondence related observations form the motivaton of the
current proposal.

In this communication we recycle Cruz et al.'s \cite{2} work on the
classical and quantum position-dependent mass oscillators. To elaborate on
the ordering ambiguity, we properly amend some of the results reported in
their work and discuss the classical-quantum mechanical correspondence for
the PDM harmonic oscillators. In section II, we recollect the quantum
position-dependent mass (PDM) oscillator \cite{2}. We emphasis that the
quantum harmonic oscillator's creation $\hat{A}^{+}$ and annihilation\ $\hat{%
A}^{-}$ operators not only satisfy the commutation relation $[\hat{A}^{-},%
\hat{A}^{+}]=1$ (as in \cite{2}), but also they satisfy the well known
complementary textbook condition $\hat{H}=\hat{A}^{+}\hat{A}^{-}+1/2=\hat{A}%
^{-}\hat{A}^{+}-1/2$. Such a condition results a unique quantum PDM
oscillator Hamiltonian, unlike the two partner-like oscillator Hamiltonians
reported by Cruz et al.\cite{2}. In section III, we use a point canonical
transformation (PCT) and show that such a unique quantum PDM oscillator
Hamiltonian represents the quantum mechanical correspondence of the
classical PDM oscillator Hamiltonian. In connection with the
ordering-ambiguity associated with the PDM von Roos Hamiltonian (1) \cite{1}%
, we observe that one "\textit{unique"} ordering-ambiguity parametric set $%
j=l=-1/4$ and $k=-1/2$ is obtained (namely, that of Mustafa and
Mazharimousavi's \cite{4}). To show that such a parametric set is not just a
manifestation of the quantum PDM oscillator Hamiltonian, we consider (in
section VI) the classical and quantum mechanical correspondence for
quasi-free PDM particles moving under the influence of their own PDM force
fields. We conclude in section V.

\section{Quantum position-dependent mass harmonic oscillator}

In their general considerations of the quantum position-dependent mass
oscillator, Cruz et al. \cite{2} have used a PDM Hamiltonian of the form%
\begin{equation}
\hat{H}=-\frac{1}{2}m\left( x\right) ^{\boldsymbol{a}}\partial _{x}m\left(
x\right) ^{2b}\partial _{x}m\left( x\right) ^{a}+V\left( x\right) \text{ ; \ 
}a+b=-\frac{1}{2}.
\end{equation}%
Which is in fact obtained from the PDM von Roos Hamiltonian (1) using the
continuity conditions at the abrupt heterojunction $j=l$, where $j=l=a$ and $%
k=2b$ are the parametric mappings between Hamiltonian (1) and (2). Their
construction of the harmonic oscillator creation%
\begin{equation}
\hat{A}^{+}=-\frac{1}{\sqrt{2}}m^{a}\partial _{x}m^{b}+W_{a}\left( x\right) ,
\end{equation}%
and annihilation%
\begin{equation}
\hat{A}^{-}=\frac{1}{\sqrt{2}}m^{b}\partial _{x}m^{a}+W_{a}\left( x\right) ,
\end{equation}%
operators have led them to two Hamiltonians%
\begin{equation}
\hat{H}^{+}=\hat{A}^{+}\hat{A}^{-}=-\frac{1}{2}m^{a}\partial
_{x}m^{2b}\partial _{x}m^{a}+V_{a}^{+}\left( x\right) =\hat{T}%
_{a}^{+}+V_{a}^{+}\left( x\right) ,
\end{equation}%
(i.e., equation (25) of \cite{2}) and%
\begin{equation}
\hat{H}^{-}=\hat{A}^{-}\hat{A}^{+}=-\frac{1}{2}m^{b}\partial
_{x}m^{2a}\partial _{x}m^{b}+V_{a}^{-}\left( x\right) =\hat{T}%
_{a}^{-}+V_{a}^{-}\left( x\right) ,
\end{equation}%
(i.e., equation (26) of \cite{2}). Where their $V_{a}^{+}\left( x\right) $
and $V_{a}^{-}\left( x\right) $ are defined in equations (27) and (28) of
their paper, respectively. At this point, the last term in their expression
for $V_{a}^{-}\left( x\right) $ in (28) should be removed (no such term
should be there) and consequently their corresponding $V_{a}^{\pm }\left(
x\right) $ in (30) should read%
\begin{equation}
V_{a}^{\pm }\left( x\right) =\frac{1}{2}\left( \int\limits^{x}\sqrt{m\left(
u\right) }du\right) ^{2}-\frac{\left( 4a+1\right) ^{2}}{8}\left[ \left( 
\frac{1}{\sqrt{m}}\right) ^{^{\prime }}\right] ^{2}\mp \frac{\left(
4a+1\right) }{4}\frac{1}{\sqrt{m}}\left( \frac{1}{\sqrt{m}}\right)
^{^{\prime \prime }}\mp \frac{1}{2}.
\end{equation}%
Obviously, our third term in (7) does not agree with their second term in
(30).

Indeed one would use the oscillator commutation relation $[\hat{A}^{-},\hat{A%
}^{+}]=1$ to obtain $W_{a}\left( x\right) $ given in their equation (29).
However, the harmonic oscillator is also well known to have one unique
Hamiltonian given by the textbook complementary relation%
\begin{equation}
\hat{H}=\hat{A}^{+}\hat{A}^{-}+\frac{1}{2}=\hat{A}^{-}\hat{A}^{+}-\frac{1}{2}%
.
\end{equation}%
Then the two Hamiltonians in (5) and (6) are not two different Hamiltonians
but they represent the constituents of one Hamiltonian $\hat{H}=\hat{H}^{\pm
}\pm 1/2$. This would, in turn, suggest that the corresponding kinetic
energy operators satisfy the relation%
\begin{equation}
\hat{T}_{a}^{+}=\hat{T}_{a}^{-}\Longrightarrow -\frac{1}{2}m^{a}\partial
_{x}m^{2b}\partial _{x}m^{a}=-\frac{1}{2}m^{b}\partial _{x}m^{2a}\partial
_{x}m^{b}.
\end{equation}%
Clearly, this relation can only be satisfied if and only if $a=b$. This
result, when substituted in the corresponding von Roos constraint $a+b=-1/2$%
, would immediately imply that $a=b=-1/4$. \ Consequently, the quantum
harmonic oscillator PDM-Hamiltonian in (8) is unique and represented by%
\begin{equation}
\hat{H}=-\frac{1}{2}\frac{1}{\sqrt[4]{m}}\partial _{x}\frac{1}{\sqrt{m}}%
\partial _{x}\frac{1}{\sqrt[4]{m}}+\frac{1}{2}\left( \int\limits^{x}\sqrt{%
m\left( u\right) }du\right) ^{2}.
\end{equation}%
This result not only shows that one unique quantum PDM oscillator's
Hamiltonian is obtained but also exactly represents the quantum mechanical
correspondence of the classical PDM harmonic oscillator Hamiltonian. We
discuss this in the following section.

\section{Harmonic oscillator classical and quantum mechanical
correspondence: point canonical transformation}

Let us consider a classical PDM-particle moving under the influence of a
potential field $V\left( x\right) $. Then the corresponding Hamiltonian
would read%
\begin{equation}
\mathcal{H}_{x}=\frac{p_{x}^{2}}{2m\left( x\right) }+V\left( x\right) =\frac{%
1}{2}m\left( x\right) \,\dot{x}^{2}+V\left( x\right) \text{ ; \ }\dot{x}=%
\frac{dx}{dt}.
\end{equation}%
Which, under PCT%
\begin{equation}
q^{\prime }\left( x\right) =\sqrt{m\left( x\right) }\Longrightarrow q\left(
x\right) =\int\limits^{x}\sqrt{m\left( u\right) }du,
\end{equation}%
would be transformed into $\mathcal{H}_{q}$ such that%
\begin{equation}
\mathcal{H}_{q}=\frac{1}{2}\dot{q}^{2}+V\left( q\right) \text{ ; }\dot{q}=%
\frac{dq}{dt}.
\end{equation}%
Where $\mathcal{H}_{q}$ represents a classical particle with a "\textit{%
constant unit mass"} moving under the influence of a potential field $%
V\left( q\right) =V\left( q\left( x\right) \right) $ with a momentum $P_{q}=%
\dot{q}$. We may now safely recollect that $\mathcal{H}_{q}$ can be
factorized into%
\begin{equation}
\mathcal{H}_{q}=a^{-}a^{+}=a^{+}a^{-}=\frac{1}{2}P_{q}^{2}+V\left( q\right) ,
\end{equation}%
such that%
\begin{equation}
a^{\pm }=\mp i\frac{P_{q}}{\sqrt{2}}+G\left( q\right) ,
\end{equation}%
satisfy the Poisson bracket%
\begin{equation}
\left\{ a^{-},a^{+}\right\} =\frac{\partial a^{-}}{\partial P_{q}}\frac{%
\partial a^{+}}{\partial q}-\frac{\partial a^{-}}{\partial q}\frac{\partial
a^{+}}{\partial P_{q}}=i
\end{equation}%
for the harmonic oscillator with a "\textit{constant unit mass"}. This
condition on $a^{\pm }$ would yield that%
\begin{equation}
G\left( q\right) =\frac{q}{\sqrt{2}}\Longrightarrow V\left( q\right) =\frac{1%
}{2}q^{2}=\frac{1}{2}\left( \int\limits^{x}\sqrt{m\left( u\right) }du\right)
^{2}.
\end{equation}%
Of course, for a given PDM $m\left( x\right) $ one may then find the
corresponding $V\left( x\right) $. However, this readily lies far beyond our
current proposal.

The quantum mechanical correspondence of such a classical model would,
moreover, transform Hamiltonian (2), using substitution $\psi \left(
x\right) =m\left( x\right) ^{1/4}\varphi \left( q\right) $ in $\hat{H}\psi
\left( x\right) =E\psi \left( x\right) $, into%
\begin{equation}
\hat{H}_{q}=-\frac{1}{2}\partial _{q}^{2}+V_{eff}\left( q\left( x\right)
\right) =\frac{1}{2}\hat{P}_{q}^{2}+V_{eff}\left( q\right) ,
\end{equation}%
where $\hat{P}_{q}=-id/dq=-i\partial _{q}$ analogous to the linear momentum
operator (with $\hbar =1$) for a quantum particle with a \textit{constant
unit mass,} 
\begin{equation}
V_{eff}\left( q\right) =\frac{1}{8}\left( 1+4b\right) F_{1}\left( q\right) -%
\frac{1}{2}\left[ \frac{9}{16}+a\left( a+2b+1\right) +2b\right] F_{2}\left(
q\right) +V\left( q\right) ,
\end{equation}%
and%
\begin{equation}
F_{1}\left( q\right) =\frac{m^{\prime \prime }\left( x\right) }{m\left(
x\right) ^{2}}\text{ \ ; \ \ }F_{2}\left( q\right) =\frac{m^{\prime }\left(
x\right) ^{2}}{m\left( x\right) ^{3}}.
\end{equation}%
Where $F_{1}\left( q\right) $ and $F_{2}\left( q\right) $ are two smooth
functions manifestly introduced by the ordering-ambiguity in (1) .
Obviously, $V_{eff}\left( q\right) $ in (19) represents an effective
potential field produced by the PDM particle itself (represented by the
first two terms) and the interaction potential. Let us now use the harmonic
oscillator creation and annihilation operators\textit{\ }%
\begin{equation}
\hat{b}^{+}=G\left( q\right) -\frac{i\hat{P}_{q}}{\sqrt{2}}=G\left( q\right)
-\frac{1}{\sqrt{2}}\frac{d}{dq},
\end{equation}%
and%
\begin{equation}
\hat{b}^{-}=G\left( q\right) +\frac{i\hat{P}_{q}}{\sqrt{2}}=G\left( q\right)
+\frac{1}{\sqrt{2}}\frac{d}{dq},
\end{equation}%
respectively, for a quantum particle with a \textit{constant unit mass}.
These operators satisfy the commutation relation $\left[ \hat{b}^{-},\hat{b}%
^{+}\right] =1$ and imply that%
\begin{equation}
G\left( q\right) =\frac{q}{\sqrt{2}}\Longrightarrow V\left( q\right) =\frac{1%
}{2}q^{2}
\end{equation}%
Moreover, the condition $\hat{H}_{q}=\hat{b}^{+}\hat{b}^{-}+\frac{1}{2}=\hat{%
b}^{-}\hat{b}^{+}-\frac{1}{2}$ implies that%
\begin{equation*}
\hat{H}_{q}=-\frac{1}{2}\partial _{q}^{2}+G\left( q\right) ^{2}=-\frac{1}{2}%
\partial _{q}^{2}+\frac{1}{2}q^{2}.
\end{equation*}%
If the classical Hamiltonian in (14) (along with $V\left( q\right) $ of
(17)) is to find its quantum mechanical counterpart in (18) (along with $%
V_{eff}\left( q\right) $ of (19) and (23)) then the first two terms of the
effective potential in (19) should vanish identically to yield $b=-1/4$ and $%
a=-1/4$ (i.e., Mustafa and Mazharimousavi's \cite{4} ordering parametric
set). Now, using the substitution $\varphi \left( q\right) =m\left( x\right)
^{-1/4}\psi \left( x\right) $ in $\hat{H}_{q}\varphi \left( q\right)
=E\varphi \left( q\right) $, one may easily show that%
\begin{equation*}
\hat{H}_{q}=-\frac{1}{2}\partial _{q}^{2}+\frac{1}{2}q^{2}\Longrightarrow 
\hat{H}=-\frac{1}{2}\frac{1}{\sqrt[4]{m}}\partial _{x}\frac{1}{\sqrt{m}}%
\partial _{x}\frac{1}{\sqrt[4]{m}}+\frac{1}{2}\left( \int\limits^{x}\sqrt{%
m\left( u\right) }du\right) ^{2}.
\end{equation*}%
Which is indeed in exact accord with Hamiltonian (10). This result not only
shows that one unique quantum PDM oscillator's Hamiltonian is obtained but
also exactly represents the quantum mechanical correspondence of the
classical PDM harmonic oscillator Hamiltonian given in (10).

Hereby, a question of delicate nature arises in the process as to whether
such a parametric set is only associated to the PDM harmonic oscillator
problem. One would therefore invest similar procedure in a different model
to test it. This is done in the following section.

\section{Quasi-free PDM-particle; Classical and quantum correspondence}

Consider a free PDM quantum particle (i.e., $V\left( x\right) =0)$) moving
under the influence of its own PDM-field (hence quasi-free PDM-particle) 
\cite{3,7}. Using the point canonical transformation (12) along with the
substitutions $\psi \left( x\right) =m\left( x\right) ^{1/4}\varphi \left(
q\right) $ and $V\left( x\right) =0$ in $\hat{H}\psi \left( x\right) =E\psi
\left( x\right) $, then $\hat{H}$ of (2) would transform into $\hat{H}_{q}$
so that%
\begin{equation}
\hat{H}_{q}=-\frac{1}{2}\partial _{q}^{2}+V_{eff}\left( q\left( x\right)
\right) =\frac{1}{2}\hat{P}_{q}^{2}+V_{eff}\left( q\right) ,
\end{equation}%
where, 
\begin{equation}
V_{eff}\left( q\right) =V_{eff}\left( q\left( x\right) \right) =\frac{1}{8}%
\left( 1+4b\right) F_{1}\left( q\right) -\frac{1}{2}\left[ \frac{9}{16}%
+a\left( a+2b+1\right) +2b\right] F_{2}\left( q\right) ,
\end{equation}

Now consider the classical Hamiltonian for a free PDM-particle moving under
the influence of its own PDM-field%
\begin{equation}
\mathcal{H}_{x}=\frac{p_{x}^{2}}{2m\left( x\right) }=\frac{1}{2}m\left(
x\right) \,\dot{x}^{2}\text{ ; \ }\dot{x}=\frac{dx}{dt}.
\end{equation}%
Which, under the PCT would be transformed into $\mathcal{H}_{q}$ such that%
\begin{equation}
\mathcal{H}_{q}=\frac{1}{2}\dot{q}^{2}.
\end{equation}%
The transformed Hamiltonian $\mathcal{H}_{q}$ represents a free particle
with a \textit{constant unit mass} moving in $q$-space with a conserved
momentum 
\begin{equation}
P_{q}=\dot{q}\longrightarrow \dot{q}\left( x\right) =\dot{q}\left(
x_{0}\right) .
\end{equation}%
(cf., e.g., Mazharimousavi and Mustafa \cite{3} for more details on this
issue). Under such settings, we recast $\mathcal{H}_{q}$ to read%
\begin{equation}
\mathcal{H}_{q}=\frac{1}{2}P_{q}^{2}.
\end{equation}%
If the quantum mechanical Hamiltonian in (24) is to correspond to the
classical Hamiltonian in (29), then the two terms of the effective potential
in (25) should vanish identically. That is,%
\begin{equation}
\left( 1+4b\right) =0\text{ , and }\frac{9}{16}+a\left( a+2b+1\right) +2b=0,
\end{equation}%
which would, again, immediately suggest that $b=a=-1/4$ (i.e., Mustafa and
Mazharimousavi's \cite{4} ordering-ambiguity parametric set).

\section{Conclusion}

We have recycled the work of Cruz et al.'s \cite{2} on the classical and
quantum position-dependent mass oscillators. In addition to the amendments
reported above, we have discussed the classical and quantum mechanical
correspondence for the PDM harmonic oscillators to conjecture on the
feasibility of the ordering ambiguity parametrization. We have used the
creation and annihilation operators of the quantum PDM harmonic oscillator
along with the commutation relation $[\hat{A}^{-},\hat{A}^{+}]=1$ and the
complementary relation $\hat{H}=\hat{A}^{+}\hat{A}^{-}+\frac{1}{2}=\hat{A}%
^{-}\hat{A}^{+}-\frac{1}{2}$ to show that the harmonic oscillator
PDM-Hamiltonian in (10) is unique and represented by%
\begin{equation}
\hat{H}=-\frac{1}{2}\frac{1}{\sqrt[4]{m\left( x\right) }}\partial _{x}\frac{1%
}{\sqrt{m\left( x\right) }}\partial _{x}\frac{1}{\sqrt[4]{m\left( x\right) }}%
+\frac{1}{2}\left( \int\limits^{x}\sqrt{m\left( u\right) }du\right) ^{2}.
\end{equation}%
We have shown that this quantum PDM oscillator's Hamiltonian is not only
unique but also exactly represents the quantum mechanical correspondence of
the classical PDM harmonic oscillator Hamiltonian (documented in section
III). Moreover, we have reported that such unique representation of the PDM
kinetic energy term in (31) is not only a by-product of the quantum PDM
oscillator, but also a by-product of the quantum mechanical correspondence
of the classical quasi-free PDM Hamiltonians (documented in sections III and
IV). In fact, one may go further and conjecture (from both PCT-transformed
harmonic oscillator and quasi-free PDM-Hamiltonians) that such unique
representation holds true also for any potential $V\left( x\right) $ (which
would PCT-transform into $V\left( q\left( x\right) \right) =V\left( q\right) 
$ leaving the first two terms of the effective potential (19) to vanish
identically).

In short, the examples discussed above show that the classical and quantum
mechanical correspondence leaves no doubt that the kinetic energy operator
in the von Roos Hamiltonian (1) now finds its \textit{unique} representation
through Mustafa and Mazharimousavi's \cite{4} ordering-ambiguity parametric
set to read%
\begin{equation}
\hat{T}_{x}=-\frac{1}{2}\frac{1}{\sqrt[4]{m\left( x\right) }}\partial _{x}%
\frac{1}{\sqrt{m\left( x\right) }}\partial _{x}\frac{1}{\sqrt[4]{m\left(
x\right) }}.
\end{equation}%
This should very likely settle down the ordering-ambiguity conflict.


\begin{thebibliography}{99}
\bibitem{1} O. Von Roos, Phys. Rev. \textbf{B 27} (1983) 7547.

\bibitem{2} S. Cruz y Cruz, J Negro, L. M. Nieto, Phys. Lett. \textbf{A 369}
(2007) 400.

\bibitem{3} J. M. L\'{e}vy-Leblond, Phys. Rev. \textbf{A 52} (1995) 1845.

S H Mazharimousavi, O Mustafa, arXiv: 1208.1095

\bibitem{4} O Mustafa, S.Habib Mazharimousavi, Int. J. Theor. Phys \ \textbf{%
46} (2007) 1786.

R Koc, G Sahinoglu, M Koca, Eur. Phys. J. \textbf{B48 }(2005) 583.

Q. G. Zhu, H. Kroemer, Phys. Rev. \textbf{B 27} (1983) 3519.

D. J. Ben Daniel, C. B. Duke, Phys. Rev. \textbf{152} (1966) 683.

T. Gora, F. Williams, Phys. Rev. \textbf{177} (1969) 1179.

T. Li, K. J. Kuhn, Phys. Rev. \textbf{B 47} (1993) 12760.

\bibitem{5} A de Souza Dutra, C A S Almeida, Phys Lett. \textbf{A 275}
(2000) 25.

\bibitem{6} S. Cruz y Cruz, O Rosas-Ortiz, J Phys \textbf{A}: Math. Theor. 
\textbf{42} (2009) 185205.

\bibitem{7} O Mustafa, S H Mazharimousavi, Phys. Lett. \textbf{A 358} (2006)
259.

\bibitem{8} A D Alhaidari, Phys. Rev. \textbf{A 66} (2002) 042116.

\bibitem{9} O Mustafa, S H Mazharimousavi, J. Phys. \textbf{A}: Math. Gen. 
\textbf{39} (2006) 10537.

\bibitem{10} B Bagchi, A Banerjee, C Quesne, V M Tkachuk, J. \ Phys. \textbf{%
A}; Math. Gen. \textbf{38} (2005) 2929.

V. Milanovi\'{c}, Z. Ikoni\'{c}, J. Phys. \textbf{A}; Math. Gen. \textbf{32}
(1999) 7001.

B. Roy, P. Roy, Phys. Lett. \textbf{A 340 }(2005) 70.

\bibitem{11} O Mustafa, S H Mazharimousavi, Phys. Lett. \textbf{A 357}
(2006) 295.

\bibitem{12} O Mustafa, S H Mazharimousavi, J Phys \textbf{A: }Math. Theor.%
\textbf{41 (}2008\textbf{) }244020.

\bibitem{13} R Koc, G Sahinoglu, M Koca, Eur. Phys. J. \textbf{B48 }(2005)
583.

\bibitem{14} B. Bagchi, P. Gorain, C. Quesne and R. Roychoudhury, Mod. Phys.
Lett. \textbf{A19} (2004) 2765

\bibitem{15} O Mustafa, S H Mazharimousavi, Phys. Lett. \textbf{A 373}
(2009) 325.

\bibitem{16} O. Mustafa, J Phys \textbf{A: }Math. Theor.\textbf{43 (}2010%
\textbf{) }385310.

\bibitem{17} O. Mustafa, J Phys \textbf{A: }Math. Theor.\textbf{44 (}2011%
\textbf{) }355303.
\end{thebibliography}
\end{document}